\documentclass[twocolumn]{aastex61}
\pdfoutput=1 
\usepackage{amsmath,amstext}
\usepackage[T1]{fontenc}
\usepackage{apjfonts} 
\usepackage{booktabs}
\usepackage{tabularx}
\usepackage[figure,figure*]{hypcap}

\newcommand{\makeinvisible}[1]{}
\newcommand{\msolar}{M$_{\odot}$}
\newcommand{\mej}{$m_{\textrm{ej}}$}

\begin{document}

\newcommand{\nuaffil}{Center for Interdisciplinary Exploration and Research in Astrophysics (CIERA)
and
Department of Physics and Astronomy,
Northwestern University,
2145 Sheridan Road,
Evanston, IL 60208,
USA}

\title{On GW170817 and the Galactic Binary Neutron Star Population}

\author{Chris Pankow}
\affiliation{\nuaffil}

\begin{abstract}
GW170817/GRB170817A, a short gamma-ray burst arising from a low-mass compact object merger was the first multi-messenger discovery of a compact binary system outside the local galactic neighborhood. From gravitational-wave measurements, we know GW170817 has a wide range of plausible component masses, depending also on less well-constrained properties such as the spin and tidal deformability of the component stars. The kilonova light curve --- and hence the total ejecta mass from a given source --- depends on the relative contribution of dynamical ejecta and other sources such as disk winds. Electromagnetic observations and model fitting of the ejecta properties from the subsequent kilonova detection provided values of the ejecta mass from the merger. These values, when combined with the gravitational-wave measurement disfavors an equal-mass configuration, with the level of disagreement dependent on the assumed amount of ejecta mass of dynamical origin. Within the confines of our own galaxy, several binary neutron star systems along with measurements of their component masses have been made. If those distributions are indicative of a universal distribution, the joint measurement of the component masses of GW170817 represents an outlier. This tension is not easily resolvable from physical arguments, as the proposed pathways which form binary neutron stars do not often produce very asymmetrical pairs. Even accounting for the uncertainty associated with the total mass of the dynamical ejecta, this tension suggests that the distribution of binary neutron star masses in the galaxy is not indicative of those in other galaxies.
\end{abstract}

\section{Introduction} \label{sec:intro}

GW170817/GRB170817A~\citep{2017PhRvL.119p1101A}, is the closest GRB yet localized~\citep{2017ApJ...848L..13A} and the first binary neutron star (BNS) merger observed with a gravitational wave (GW) interferometer network including the Advanced LIGO~\citep{2010CQGra..27h4006H} and Advanced Virgo~\citep{2015CQGra..32b4001A} interferometers. Given its proximity, the host galaxy and post-merger electromagnetic transient (AT2017gfo) was identified~\citep{2017Sci...358.1556C,2017ApJ...848L..16S,2017ApJ...848L..24V,2017Natur.551...64A,2017ApJ...848L..27T,2017Natur.551...67P,2017ApJ...850L...1L} within about 12 hours, and intensive programs of X-ray, optical/infrared, and radio observations soon followed~\citep{2017ApJ...848L..12A}. Light curves measured over the next several weeks identified the hallmarks of a kilonova~\citep{1998ApJ...507L..59L,2017LRR....20....3M} --- emission driven by neutron-rich ejected material undergoing r-process nucleosynthesis. Depending on the opacity of the resultant ejecta, the flux from this decay-heated ejecta peaks on the $\sim$day to $\sim$week timescale~\citep{2014MNRAS.441.3444M}. 

In the case of GW170817, it is possible that several types of ejecta with differing opacities and velocities are exhibited: a fast, ``blue'', low opacity component originating from the shocked interface of the two merging NS~\citep{2013ApJ...773...78B,2016MNRAS.460.3255R}, a ``red'', higher opacity component originating from an equatorally focused ``tail'' of material dynamically ejected from the binary~\citep{2013ApJ...775...18B,2013RSPTA.37120272R,2013PhRvD..87b4001H,2016MNRAS.460.3255R}. Later evolution of an accretion disk around the remnant may also produce winds~\citep{2014MNRAS.443.3134P,2015ApJ...813....2M,2015MNRAS.450.1777K,2015MNRAS.446..750F,2017ApJ...846..114F}, resulting in contributions to both blue and red emission.

Studies of spectra and photometry~\citep{2017ApJ...848L..17C,2017ApJ...848L..18N,2017ApJ...848L..19C,2017Natur.551...80K,2017Natur.551...75S} characterized the kilonova emission properties, with joint models parameterzing them primarily by the mass (\mej), opacity, and velocity of the ejected material~\citep{2017CQGra..34j5014D,2017ApJ...849...12C}. More complicated models~\citep{2017ApJ...850L..19M,2018ApJ...856..101M,2018ApJ...861L..12L,2018ApJ...855..103P,2018Natur.554..207M,2018MNRAS.474L..12C,2018ApJ...861L..12L,2018arXiv180509371C} have been proposed to explain features in the light curves which are not easily explained by a two-component model. Some of these studies examine a detailed geometric~\citep{2018MNRAS.478.3298W,2017ApJ...850L..37P} parameterization of the merger and post-merger dynamics. If the ejecta originates from the merger dynamics of the system, the ejecta mass must be greater than 10$^{-2}$ \msolar in order to explain the peak and duration of the light curves in optical bands, given the distance~\citep{2017ApJ...848L..31H,2018ApJ...854L..31C} to the source.

Analysis of the gravitational-wave data using Markov chain Monte Carlo (MCMC) sampling~\citep{2010PhRvD..81f2003V,2015PhRvD..91d2003V,2017PhRvL.119p1101A,2018arXiv180511579T} with putative gravitational waveforms provided a Bayesian posterior measurement of the physical parameters of the pre-merger binary, including component masses and NS spin information. Those waveforms also include a set of finite-size effects~\citep{2008ApJ...677.1216H,2013PhRvD..88d4042R} which encode some the effects of the tidal interactions on the bodies in the binary, and in turn dependent on the equation of state (EoS) of the nuclear matter composing the NS. These analyses are able to measure the chirp mass of the binary very precisely~\citep{2005PhRvD..71h4008A}, but leave a wide degeneracy in the component masses which extends from an equal-mass binary (1.37 + 1.37) into values of mass ratio $q=m_2/m_1 \sim 0.7 - 0.4$ (depending on the allowed component spins) and consequently produce a binary nominally containing a very massive primary, likely a black hole.

\citet{2012ApJ...757...55O}, using six radio-detected system with pulsars and close orbits, carried out a fit of the neutron star masses to a normal distribution, obtaining a narrow set of Bayesian confidence contours around $\mu=1.33$ and $\sigma=0.09$~\citep{2016ARA&A..54..401O}. The equal-mass configurations fall comfortably within the observed Milky Way population of binary neutron stars. It is pertinent to note that the equal-mass configurations supported by the GW data intersect almost exactly with the Galactic distribution near its mean of 1.33 \msolar. However, the peak of the GW primary/secondary mass posteriors in \citet{2017PhRvL.119p1101A} and \cite{2018arXiv180511579T} is not at equal-mass, instead peaking near $q \sim 0.7 - 0.8$, implying that significant support from the gravitational-wave data exists for a BNS configuration which is not consistent with the inferred Galactic distribution. Moreover, the results of several simulations~\citep{2016CQGra..33r4002L,2017PhRvD..95b4029D,2017PhRvD..95d4045D} of binary neutron star mergers and their ejecta products show a strong dependence of the ejecta mass on the mass ratio of the binary. The ejecta mass scales with increasingly asymmetric binary configurations. Projecting this against the gravitational-wave data implies that equal-mass configurations would not have produced the emission observed because the brightness of the light curves are indicative of a high ejecta mass and thus asymmetric mass ratio. It is this tension we quantify and explore here.

\section{Joint Mass Measurements} \label{sec:joint}

Here, we represent the joint measurement as a posterior distribution, treating the EM data as a likelihood, with the ejecta mass constraints projected into the GW parameter space via a fit, and the GW measurement applied as a prior distribution. The gravitational-wave measurements reported in~\cite{2018arXiv180511579T}~\footnote{The samples used in this work can be found at \url{https://dcc.ligo.org/LIGO-P1800061/public}} provide a distribution on the component masses, which will be denoted as $p(m_1, m_2|\textrm{GW})$. Two distributions on the component masses were reported in \cite{2018arXiv180511579T}, one corresponding to dimensionless component spins up to near maximal ($\chi<0.89$) and one corresponding to small spins, up to the largest measured NS spin in an electromagnetically observed BNS ($\chi<0.05$). The former set is referred to as the `high-spin' prior set and the latter as the `low-spin' prior set.

Using a fit derived from several numerical relativity simulations~\citet{2017PhRvD..95d4045D}, allows a projection of the EM ejecta distribution into the component mass space, labeled as $p(m_1, m_2|\textrm{EM})$. Several photometric measurements (see~\cite{2017ApJ...851L..21V,2018ApJ...855L..23A} for an extensive list and comparison) of the properties of the ejecta driving the kilonova have been made, most being in broad agreement with each other. The values from the two-component model in \citet{2017ApJ...848L..17C} are adopted here\footnote{Note that a three-component model is favored slightly more statistically, allowing for a moderate opacity ``purple'' ejecta component. In this model, the ejecta is split between the purple and red, a distinction we will not follow upon.}, and are modeled as a Gaussian distributed value of \mej~centered on the median and standard deviation derived from either of the 90\% intervals of $3.6^{+0.2}_{-0.1}\times 10^{-2}$ \msolar (red) and $1.4^{+0.2}_{-0.1}\times 10^{-2}$ \msolar (blue) quoted in that publication. The fitted velocities of either component ($v_{\textrm{red}}=0.15c$ and $v_{\textrm{blue}}=0.3c$) together with NS merger simulations hint that the blue ejecta originates from the shocked NS interface at merger, and the red component is dominated by a post merger disk wind~\citep{2017PhRvL.119w1102S}. Without definitive evidence to either scenario, and allowing for the possibility that either component can be contributed to in some measure by dynamical or disk winds, we consider the two distributions as bracketing the possible resulting mass distributions.

\section{Comparison with Galactic Distributions} \label{sec:joint_comp}

The distribution of NS masses has been fit to a Gaussian distribution with a mean of 1.33 \msolar~and standard deviation of 0.09 \msolar, from known binary neutron stars which will merge within a Hubble time in the local Galactic neighborhood~\citep{2012ApJ...757...55O,2016ARA&A..54..401O}. Since the fit was performed, additional systems have been discovered (a few can be found in~\cite{2017ApJ...846..170T}). J0509+3801~\citep{2018ApJ...859...93L}, is a highly eccentric, roughly equal mass binary ($q\sim0.93$). Another recent discovery is J1946+2052~\citep{2018ApJ...854L..22S}, a very tight binary with a total mass of only 2.5 \msolar. J1411+2551~\citep{2017ApJ...851L..29M} also has a comparatively low total mass, and the component masses values are not well constrained, allowing for a mass ratio of $> 0.57$. However, the equal-mass configuration of the same total mass is two NS of 1.26 \msolar, reasonably contained within the inferred Galactic distribution. Moreover, this binary is not expected to merge in a Hubble time. J1757-1854~\citep{2018MNRAS.475L..57C} is another eccentric BNS, with very well determined masses near the mean of 1.33 \msolar. J1913+1102~\citep{2016ApJ...831..150L} has similar limits on the mass ratio, but a heavier total mass, giving an equal-mass configuration of 1.44 \msolar, also comfortably within the bounds of the distribution. In \citet{2012ApJ...757...55O}, the fit does contain systems with only bounds from the mass function, but is strongly influenced by systems where the individual masses are measured precisely. If the fit were redone with the more recent discoveries, it is plausible the the inferred parameters of the distribution would not be radically changed. As such, this work will adopt these parameters as representative of the Galactic neutron star distribution for both components of the binary. However, see Section \ref{sec:concl} for discussion of other asymmetric systems in this context.

\subsection{Reweighting the Gravitational-wave Data}

In order to quantify scenarios which involve the hypothesis that GW170817 arose from a population which is modeled by a Gaussian distribution, first it is required to produce a GW sample set which is representative of the prior belief of a Gaussian distribution as opposed to the uniform. The gravitational-wave posterior samples assume a uniform prior\footnote{Due to technical reasons, while this prior is uniform, it has an irregular perimeter. The only effect of this is a possible change in the normalization constant applied to the prior.}. The prior area (here denoted as $\Omega$) covered by $p_U$ does not totally enclose $p_G$, due to the irregular boundaries. However, the strongly peaked likelihood is bounded away from the uncovered region, making biases from ignoring this part of the space unlikely. The form of the likelihood $p(d|m_1, m_2, \textrm{GW})$ is unchanged with the change of hypothesis in the prior. Thus, instead of rerunning the full parameter estimation pipeline and obtaining a new set of samples, it should be sufficient to instead resample the GW posterior distribution. The reweighting applied is:

\begin{eqnarray} \label{eqn:reweight}
p(m_1, m_2 \vert \textrm{GW}, G) & = & p_G(m_1, m_2|\mu, \sigma) / p_U(m_1, m_2|\Omega) \nonumber \\
& & \times p(m_1, m_2 \vert \textrm{GW}, U)
\end{eqnarray}

\noindent In equation \ref{eqn:reweight}, the subscript on the prior definition indicates $U$ for uniform and $G$ for Gaussian, intended to distinguish the different form of the priors. 

\begin{figure}[htp]
\includegraphics[width=\columnwidth]{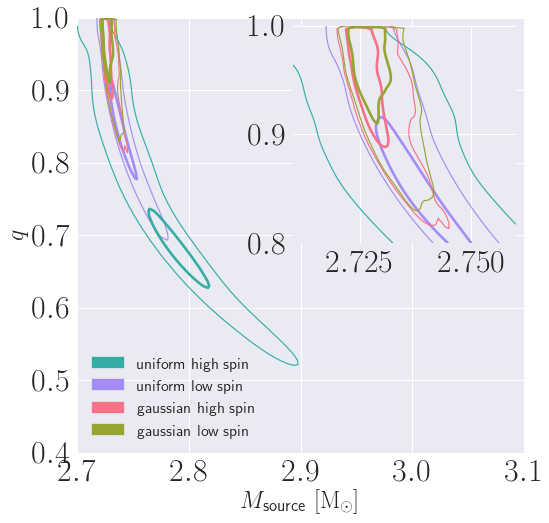}
\caption{\label{fig:distr} Comparison of the gravitational-wave high and low-spin posterior (green, purple) and their reweighted versions (red, gold). The thicker/thinner contour corresponds to the 50\%/90\% credible regions respectively. The inset in the upper right shows a zoom in on the region of interest for the reweighted distributions.}
\end{figure}

Figure \ref{fig:distr} shows the original posterior distributions overlaid with the reweighted distributions. The reweighted distribution represents the gravitational-wave posteriors under the prior assumption that the component masses are drawn from a Gaussian distribution with $\mu=1.33$ and $\sigma=0.09$.

The high-spin priors extend out to extremely high masses (well above 3 \msolar), so the reweighting procedure introduces a sharper fall off in the mass ratio, restricting the 90\% region to less than 0.7 and the total mass to less than 2.9 \msolar. The effect on the low-spin prior is less pronounced, since it did not extend beyond this region to begin with.

\subsection{Application of the Electromagnetic Constraints}

The likelihood defined in section \ref{sec:joint}, confines the most likely values of either component's ejecta mass to narrow ranges. We model this distribution in the ejected matter with a Gaussian distribution centered on the reported median value, and a standard deviation which gives 90\% credible intervals consistent with that measurement. The actual shape of that distribution is mostly unimportant, given how narrow the credible region is. Since it is formulated in the ejecta mass, the constraint must be projected into the component mass plane so we can apply the gravitational-wave prior. We accomplish this using the fitting function from \citet{2017CQGra..34j5014D}. This function is derived from a set of binary neutron star merger simulations with a variety of configurations in the component masses and the equation of state (EoS). This introduces an additional dependence on the NS radii from the EoS through its implicit radius-mass relationship.

We seek to apply the EM constraints in the component mass space by use of the fits to infer the distribution of $m_1, m_2$ from a given draw from the ejecta mass distribution. However, a confounding factor in this analysis is that the ejecta fits come with a 72\% relative error in the fitted ejecta mass for a given value of $m_1, m_2$ and corresponding NS radii. The overall effect of this error would be to widen the distribution of the component masses obtained from a distribution of the ejecta mass. In \citet{2017ApJ...850L..39A}, a log-normal distribution centered on the ejecta mass sample was employed with a spread parameter which produced the desired error in the ejecta distribution. We adopt this distribution here as well, and incorporate the uncertainty introduced by sampling from the joint component mass distribution in a fashion consistent with both the mass posteriors and the log-normal spread in the ejecta distribution with a MCMC. The MCMC is then sampling the component mass space using $p(m_1, m_2|\textrm{GW},M)$ as a proposal, and only accepting jumps when they are consistent with both the lognormal spread distribution and $p(m_1, m_2|\textrm{EM})$.

Since there are two competing hypotheses (the NS masses follow a uniform distribution versus the Galactic NS distribution), a natural way to quantify the preference for a given model versus another is the Bayes factor (BF), the ratio of the evidences under either hypothesis. The evidence --- the quantity which normalizes the posterior in Bayes' rule --- is calculated as:

\begin{equation} \label{eqn:evidence}
Z_M = \int dm_1dm_2 p(m_1, m_2|\textrm{GW},M) p(m_1, m_2|\textrm{EM})
\end{equation}

\noindent Wherein above and throughout the remainder of the work, $M$ refers to the model: $G$ for Gaussian reweighted priors and $U$ for uniform priors. For each EoS, the evidence is calculated from importance sampling each posterior.

To incorporate the EoS with the gravitational-wave data in a precise manner is difficult --- specific EoS would inscribe trajectories through a two-dimensional posterior over the tidal deformability parameters. A careful accounting is required to treat the mass and tidal distributions self-consistently. Instead, the tidal deformability is ignored in the following, its dependence will only manifest when referring to of the effect of the component radius on the ejecta fits. For the most part, the choice of EoS is unimportant in estimating the evidence for a given hypothesis. However, the EoS does play one, arguably important, role in shaping the joint observational posterior distribution: the maximum allowed NS mass. The highly asymmetric mass configurations allowed by the GW data extend to very massive primary component masses --- in some cases, beyond those even allowed by causality~\citep{1996ApJ...470L..61K} and thus implying the existence of a black hole. However, adopting an EoS imposes a maximum mass cutoff. In many cases --- particularly for the high-spin gravitational-wave posterior --- this cutoff can cut directly into the support of the likelihood, and becomes a point of departure for evidences between them.

\section{Results} \label{sec:results}

\begin{figure*}[htp]
\includegraphics[width=\columnwidth]{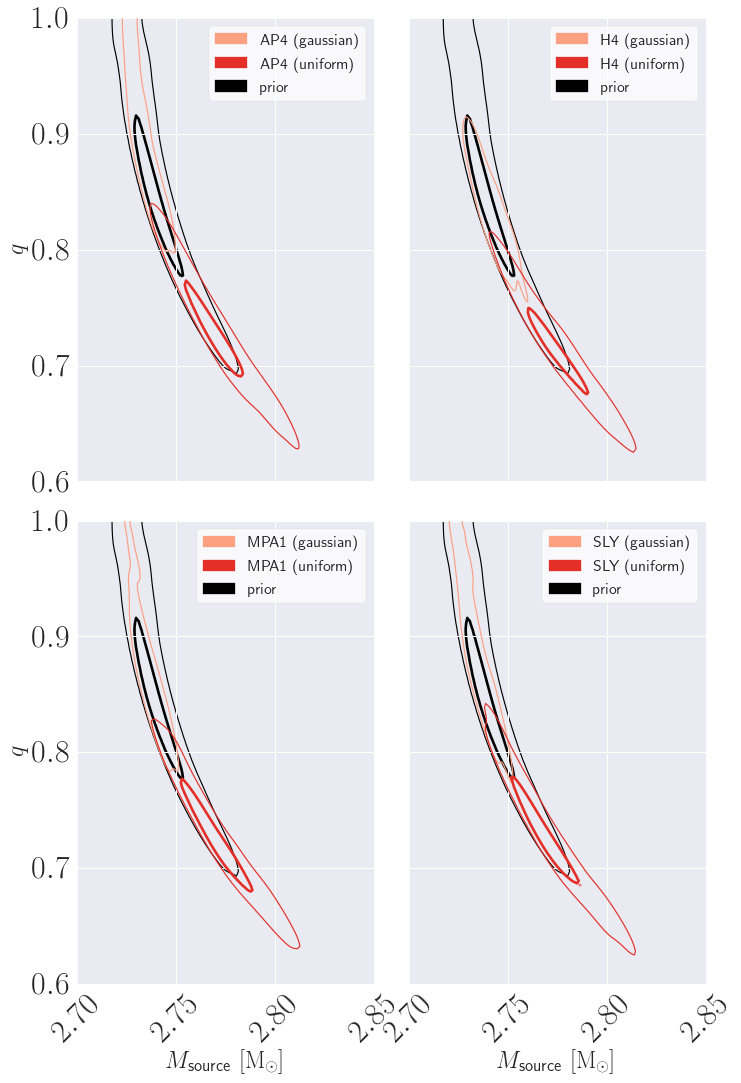}
\includegraphics[width=\columnwidth]{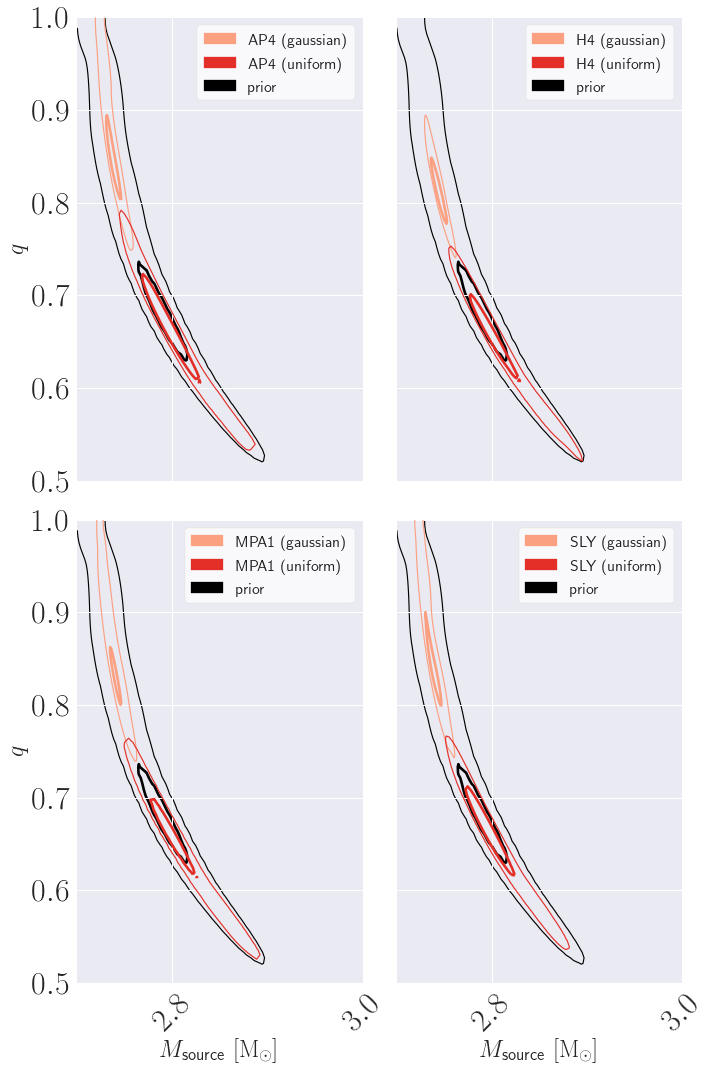}
\caption{\label{fig:mej_cons_red} Displayed here (left: low-spin, right: high-spin) are a selection of $M_{\textrm{source}}$ and $q$ distributions after applying the electromagnetic constraint through the measured ejecta mass distribution. The entirety of the ejecta associated with the red component is assumed. The light red contours (heavy/light corresponding 50\%/90\% as before) correspond to the constrained distribution under the Gaussian prior reweighting, and the dark red to the constrained distribution without the reweighting. These two contour sets should be compared to $p(M, q\vert \textrm{GW}, U)$, which is plotted in black and labeled as 'prior'. Each panel contains a different EoS, which affects the component radius and hence the inferred ejecta.}
\end{figure*}

\begin{figure*}[htp]
\includegraphics[width=\columnwidth]{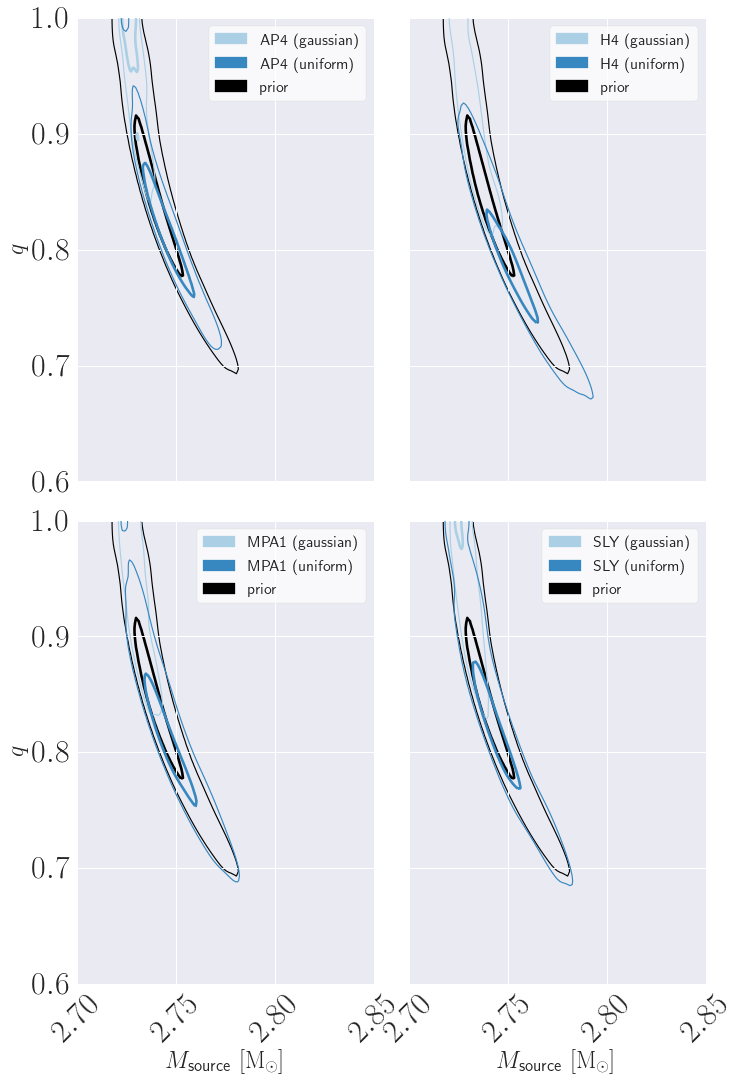}
\includegraphics[width=\columnwidth]{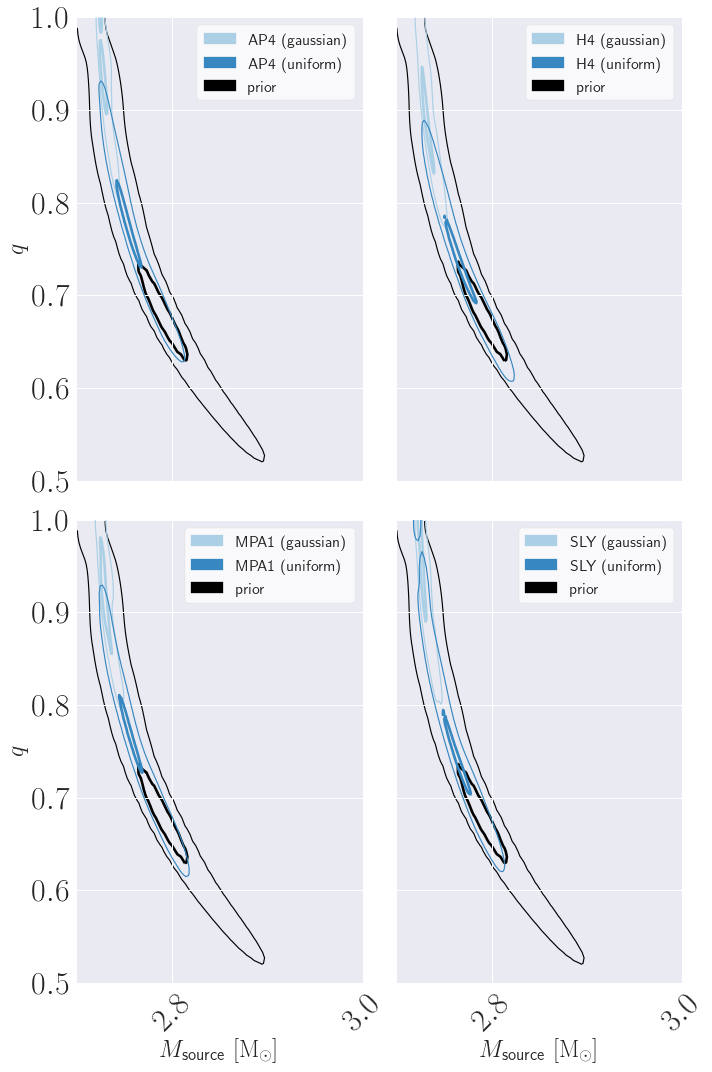}
\caption{\label{fig:mej_cons_blue} Same as Figure \ref{fig:mej_cons_red}, but assuming the ejecta mass associated with the blue kilonova component is the source of the dynamical ejecta. Light blue is Gaussian prior, and dark blue is uniform.}
\end{figure*}

In Figures \ref{fig:mej_cons_red} and \ref{fig:mej_cons_blue}, the constrained posteriors, presented in total mass (source frame) $M_{\textrm{source}}$ and mass ratio $q$ for each EoS are shown. We opt to show $M_{\textrm{source}}$ and $q$ rather than component masses, because the highly covariant and narrow posteriors in the component masses make it more difficult to discern differences. Figure \ref{fig:mej_cons_red} are the resultant posteriors if the $p(m_1, m_2|\textrm{EM})$ is a Gaussian formed from the parameters inferred from the red kilonova component and Figure \ref{fig:mej_cons_blue} is the result if the only the blue component is used. The original gravitational-wave-only credible regions are shown in black; they are different only between their allowed NS spins. The unweighted EM constrained posterior distribution is shown in blue and the Gaussian-reweighted and EM-constrained distribution is shown in red. 

Once distinguished between the red and blue ejecta model or the allowed spin magnitudes, the EM-constrained but uniformly weighted distributions occupy very similar credible regions for all the EoS considered. There is some variation in shape and size of credible regions for the Gaussian reweighted and EM constrained distributions, regardless of the spin prior applied. Comparing against their counterparts which have not been reweighted shows distinct shifts relative to the unweighted distribution in higher total mass and more asymmetric component mass combinations (e.g. $q < 1$). Broadly, comparing blue to red bears out the expectation that smaller ejecta mass values favor more equal-mass configurations. The discriminator of requiring consistency with the red ejecta mass distribution (Figure \ref{fig:mej_cons_red}) has reduced support in $q > 0.9$ for the Gaussian reweighted and low-spin case, and $q > 0.8$ for the unweighted distribution in most cases. 

Also notable is the low-spin, blue ejecta case (Figure \ref{fig:mej_cons_blue}, left panels), where the Gaussian reweighted and unweighted distributions are separated but still wholly contained within the space of the prior. This implies that low-spin prior allows for the restriction applied from the electromagnetic data as well as encompassing both the weighted and unweighted distributions. Thus, the EM data is not informative enough to induce large differences in the resulting distributions, and the evidences (equation \ref{eqn:evidence} and BF will be dominated by the Gaussian reweighting versus unweighted.

Conversely, when considering the red ejecta cases, it is likely that electromagnetic constraints are reshaping the distributions and equal-mass regions are being populated primarily by the allowed error on the fitting function itself. If the relative error were reduced, it is likely that there would be a systematic pile up of the distributions near $q \sim 0.85$ where the values of the mass ratio would be consistent with both a sharply falling support from the Galactic distribution and a sharp increase in support from the allowed ejecta values. The high-spin priors, with a greater support within the region of high electromagnetic probability, are shifted to even more asymmetric mass ratios, typically $0.6 < q < 0.65$, as well as correspondingly heavier total masses (2.85 \msolar) as opposed to their Gaussian reweighted counterparts near 2.8 \msolar.

\setlength{\tabcolsep}{0.5em}

\begin{table*}
\begin{tabularx}{0.98\textwidth}{r|ccc|ccc|ccc|ccc}
 & \multicolumn{6}{c|}{red} & \multicolumn{6}{c}{blue} \\
\hline
 & \multicolumn{3}{c|}{low spin $\chi < 0.05$} & \multicolumn{3}{c|}{high spin $\chi < 0.89$} & \multicolumn{3}{c|}{low spin $\chi < 0.05$} & \multicolumn{3}{c}{high spin $\chi < 0.89$} \\
 EOS & $\log Z_U$ & $\log Z_G$ & $\log Z_U / Z_G$ & $\log Z_U$ & $\log Z_G$ & $\log Z_U / Z_G$ & $\log Z_U$ & $\log Z_G$ & $\log Z_U / Z_G$ & $\log Z_U$ & $\log Z_G$ & $\log Z_U / Z_G$ \\
\hline
\hline
ALF2 & -9.2 & -62.3 & 53.1 & 1.8 & -43.5 & 45.3 & 	2.4 & -2.6 & 5.0 & 3.4 & -2.1 & 5.5 \\ 
AP3 & -8.0 & -58.5 & 50.5 & 1.8 & -40.6 & 42.5 & 	2.5 & -2.5 & 5.0 & 3.4 & -1.9 & 5.3 \\ 
AP4 & -7.3 & -56.5 & 49.2 & 1.8 & -38.7 & 40.6 & 	2.6 & -2.2 & 4.8 & 3.4 & -1.7 & 5.1 \\ 
ENG & -8.0 & -58.6 & 50.6 & 1.8 & -40.6 & 42.4 & 	2.8 & -1.6 & 4.4 & 3.4 & -1.0 & 4.4 \\ 
GNH3 & -14.3 & -74.5 & 60.2 & 1.7 & -54.1 & 55.7 & 	3.0 & -1.2 & 4.2 & 3.3 & -0.6 & 4.0 \\ 
H4 & -10.5 & -65.6 & 55.1 & 1.8 & -46.4 & 48.2 & 	3.1 & -0.8 & 3.9 & 3.3 & -0.3 & 3.6 \\ 
MPA1 & -8.4 & -60.0 & 51.5 & 1.8 & -41.6 & 43.4 & 	3.2 & -0.4 & 3.6 & 3.3 & 0.1 & 3.2 \\ 
MS1 & -13.6 & -73.2 & 59.6 & 1.7 & -52.9 & 54.6 & 	3.2 & -0.6 & 3.7 & 3.3 & -0.1 & 3.4 \\ 
MS1B & -12.5 & -70.7 & 58.2 & 1.7 & -50.8 & 52.5 & 	3.2 & -0.6 & 3.8 & 3.3 & -0.1 & 3.4 \\ 
SLY & -7.6 & -57.6 & 49.9 & 1.8 & -39.8 & 41.6 & 	3.3 & -0.1 & 3.4 & 3.3 & 0.4 & 2.9 \\ 
SQM3 & -6.2 & -53.9 & 47.7 & 1.8 & -36.3 & 38.2 & 	3.3 & -0.2 & 3.5 & 3.3 & 0.2 & 3.1 \\ 
WFF1 & -6.4 & -54.3 & 47.8 & 1.8 & -36.8 & 38.6 & 	3.4 & 0.2 & 3.2 & 3.3 & 0.6 & 2.7 \\ 
WFF2 & -7.0 & -55.7 & 48.7 & 1.8 & -38.1 & 39.9 & 	3.4 & 0.2 & 3.2 & 3.3 & 0.6 & 2.7 \\ 
\hline
\end{tabularx}

\caption{\label{tbl:bf_ug} Bayes factors for uniform versus Gaussian, listed for the 'red' component ejecta (left columns) and 'blue' (right columns). Each set is further subdivided into low spin and high spin priors, and for each EoS, color, and spin combination, their evidences $Z_U$ and $Z_G$, as well as the log Bayes factor.}

\end{table*}

Table \ref{tbl:bf_ug} enumerates the evidences for each model and their log ratio (the log BF) with the left set of columns derived from the low-spin results and the the right set corresponding to high-spin. The individual $Z$ columns are Equation \ref{eqn:evidence} applied to integrate the blue and red distributions in Figures \ref{fig:mej_cons_red} and \ref{fig:mej_cons_blue} and obtain $Z_U$ and $Z_G$, respectively. The log of the ratio of these two is the log BF, and this is quoted in the rightmost column per spin prior set. Though a full suite of EoS were examined, Table \ref{tbl:bf_ug} only list BF for those whose maximum masses are supported by the observation of the heaviest known NS~\citep{2010Natur.467.1081D,2013Sci...340..448A}. As expected, the BF strongly favors the uniform against the Gaussian prior for all EoS, moderated primarily by the assumed spin prior and the allowed ejecta masses. There is some moderate variation in the evidence calculated on per-EoS basis regardless of the spin prior applied. The variations in the high-spin case, however are driven by $Z_G$ while $Z_U$ is remarkably constant across most EoS.

For those EoS which are supported by current observations, there is little variation within the EoS. A pairwise comparison of evidences between EoS shows no significant preference for a given EoS particularly with the high-spin prior set. This is perhaps not surprising since we have not given special treatment to an EoS specific hypothesis, e.g. expanding the posterior distributions to include information about the tidal deformations.

\section{Discussion and Conclusions} \label{sec:concl}

We have presented Bayes factors for uniform (uniformed distribution) versus the Gaussian (Galactic distribution) hypothesis for both ``blue'' and ``red'' results, with the assumption that the entirety of the ejecta mass in either component is derived from dynamical processes (namely the polar or tidal tail). The fit connecting the gravitational wave posterior distributions to the inferred ejecta mass used in this work is calibrated primarily to the dynamically-driven ejecta. Given the rapid velocity ($v=0.3c$) associated with the blue component, it is unlikely that disk winds would produce this component, and hence we establish this as a lower bound to the resultant evidences and BFs. The corresponding upper limits assume that the dynamical tidal tails dominate the disk winds~\citep{2017PhRvL.119w1102S} and thus forms the primary source of the ejecta associated with the red component. If the red component has a significant contribution from disk winds, as is expected from its lower velocity, then the truth will lie somewhere between the two scenarios. The fitted EM velocity ($v \sim 0.15$c) of the tidally-torqued ejecta is somewhat lower than expectations from modeling (see \cite{2017CQGra..34j5014D} and tables within), which usually produces a tail of $>0.2$c. However, even the disk mass available for such a process is \emph{also} dependent on the mass ratio of the system, with asymmetric configurations promoting a larger disk~\citep{2006PhRvD..73f4027S}. While the fit used in this work does not capture these effects, the overall scaling of disk mass with mass ratio is likely also indicative of asymmetric binaries.

At the most extreme, the mass ratios obtained from the joint distribution with high-spins and red ejecta imply $q\sim 0.65$. Similar constraints on the mass ratio have been obtained with earlier work~\citep{2017ApJ...851L..45G} using different observations and methods. For the total masses allowed near this value, we obtain $m_1 \sim 1.7$ \msolar and $m_2 \sim 1.1$ \msolar. Numerical models of stellar collapse predict strongly peaked mass functions, in concordance with the observational result and fit. However, while \citet{2001ApJ...554..548F} do obtain the peak at $\sim 1.2$, they also obtain a lower probability, near-continuous tail above $\sim 1.6$. They adopt a fiducial fraction of energy driving the supernova and resultant fallback, which can enhance the birth mass of the NS. This may allow for a more massive neutron star, but it lends no support to the smaller secondary companion at asymmetric mass ratios. A categorical study of the various supernova mechanisms and progenitor types \citep{2012ApJ...749...91F} seems to lead to similar conclusions, electron capture supernova and lighter progenitors produce remnant (baryonic) masses around 1.28 \msolar. Heavier NS are produced for progenitors between 10-30 \msolar, but uncertainties in the explosion mechanism and variation in metallicity allows for a wide variety of NS masses.

However, a mass of 1.1 \msolar is unusual regardless of the prescriptions adopted in any of those simulations. Later studies by \citet{2018ApJ...856...63F} reinforce this result, producing NS between $1.2-1.7$ \msolar. It is notable, however, that while feasible within these models, the masses here would be near the outlying extremes of these results as well. One possibility is a relatively light Fe-core participating in an ultra-stripped supernova explosion~\citep{2015MNRAS.451.2123T}. However, current simulations have only produced heavier cores~\cite{2018MNRAS.479.3675M}. If GW170817 truly contains a 1.1 \msolar secondary --- further implying a heavy primary --- then this could hint towards aspects of the system's evolutionary path including significant mass transfer from the secondary star, completely stripping its envelope. Such `recycling'~\citep{2016MmSAI..87..517T} could also spin up the heavier primary --- the high-spin prior supports such a case with higher masses when allowing for faster spin. However it should be cautioned that these could also arise from the inherent degeneracies of the property estimation from gravitational-wave data. Regardless, large asymmetries would favor larger values of the ``effective spin''~\citep{2018arXiv180511579T} parameter, a mass weighted spin average --- this could further imply that one or both of the components had spin near or faster than those observed in the galaxy.

Some population synthesis, such as \cite{2013ApJ...779...72D} form binary neutron stars, incorporating prescriptions from \cite{2012ApJ...749...91F}, seem to be able to produce some of the more asymmetric configurations favored here, albeit at a lower frequency than the well established peak. It is notable that they obtain a double peaked structure when the primary and secondary distributions are separated. More recent systematic studies~\citep{2018arXiv180105433K} of double neutron star formation, including the metallicity of NGC4993 effects do produce systems with similar asymmetry to the constraints obtained here (following~\cite{1996ApJ...457..834T}).

The dominant contributor to the spread in the component mass distribution is the relative error in the ejecta fits, and we have incorporated their effects into the jointly constrained distribution. Even with 72\% relative errors, the constrained distribution favors a GW prior distribution without the additional Galactic distribution constraint. The fits used to infer the mass distribution from ejecta values is specifically only for non-spinning NS. Additional work exploring the effects of spins aligned with the system orbital angular momentum axis has appeared in \cite{2017PhRvD..95d4045D}. They find that NS systems with spins aligned with the orbital angular momentum can change the amount of ejecta produced --- however it is difficult to model how this would influence the result presented here since the variation over spin configurations (the number of which is representative but small) is of the order of the uncertainties in the ejecta mass.

Observationally, there is now growing support for asymmetric systems, even within the Galactic population. For instance, J0453+1559~\citep{2015ApJ...812..143M}, was measured to have a mass ratio of $q\sim0.75$ and secondary mass of 1.174 \msolar, but will not merge in a Hubble time. More recently, it has come to the author's attention that J1913+1102 may also be an asymmetric system, of the same order as J0453+1559, \emph{and} will merge in 500 Myr. If confirmed, this would imply that the Galactic distribution does contain GW170817-like systems, and that the updated distribution fitted would likely alleviate some of the tension explored here.

This result could also suggest that the Galactic binary neutron star mass distribution itself is either not universal or the form of it is not well-described by a Gaussian with the quoted parameters. The latter is difficult to resolve with any simple distributional model --- measurements of DNS masses tend to be confidently confined near the quoted peak of 1.33 \msolar due to the precision obtained from measuring post-Keplerian parameters. When only upper-limits are available, they can span a wide range, including the masses in the constrained distributions shown here. The fit from \cite{2012ApJ...757...55O} incorporates both types of systems and still obtains the narrow distribution used in this work. Thus, the idea that the Galactic distribution does not hold for other galaxies may be more plausible: NGC4993, the host of GW170817's progenitor stars, is a early-type, possibly lenticular, galaxy. These types exhibit characteristics of both spiral and elliptical galaxies. Moreover, observations of the host galaxy show evidence for a merger with another galaxy about a Gyr ago~\citep{2017ApJ...848L..22B,2017ApJ...848L..28L}. These events could have shaped a different formation environment or evolutionary path for the progenitor system~\citep{2017ApJ...850L..40A} of GW170817.

The uniform priors used in measuring the masses of GW170817 from the GW data can simultaneously support an origin from the Galactic distribution and the constraints implied by electromagnetic observations, regardless of the assumed color. When the gravitational-wave data is reweighted to reflect applying a prior for a MW like distribution, the BF relative to the uniform prior model disfavors the alternative. This result holds for both low-spin and high-spin priors and blue versus red ejecta, but the high-spin priors and red ejecta obtain evidences values and BF which are consistently larger than the low spins and blue colors. Consequently, when considering red ejecta and all reasonable equations of state tested, the confined distribution in the component masses has virtually no support in the range of the Milky Way DNS distribution. The separation is reduced in the blue ejecta case, particularly for the high-spin priors, with a corresponding reduction in the BF.

\acknowledgements
This work has benefited from discussions with Ben Nelson, Katie Breivik, Kyle Kremer, Pranab Ghosh, and Vicky Kalogera. Thanks also goes to Will Farr for invaluable advice on incorporating the error in the ejecta mass formula. The author is supported by the NSF grant PHY-1607709, and also acknowledges support by the Center for Interdisciplinary Exploration and Research in Astrophysics (CIERA). This research was supported in part through the computational resources from the Grail computing cluster --- funded through NSF Gravitational Physics --- and staff contributions provided for the Quest high performance computing facility at Northwestern University which is jointly supported by the Office of the Provost, the Office for Research, and Northwestern University Information Technology.

\bibliography{bibliography}

\end{document}